\documentclass{article}

    \PassOptionsToPackage{numbers, compress}{natbib}
    \usepackage[preprint]{neurips_2024}

\usepackage[utf8]{inputenc} % allow utf-8 input
\usepackage[T1]{fontenc}    % use 8-bit T1 fonts
\usepackage{hyperref}       % hyperlinks
\usepackage{url}            % simple URL typesetting
\usepackage{booktabs}       % professional-quality tables
\usepackage{amsfonts}       % blackboard math symbols
\usepackage{nicefrac}       % compact symbols for 1/2, etc.
\usepackage{microtype}      % microtypography
\usepackage{xcolor}         % colors
\usepackage{amssymb}
\usepackage{pifont}
\usepackage{authblk}
\usepackage{graphicx}
\title{Codec Does Matter: Exploring the Semantic Shortcoming of Codec for Audio
Language Model}

\author[1]{Zhen Ye}
\author[1]{Peiwen Sun}
\author[3]{Jiahe Lei}
\author[4]{Hongzhan Lin}
\author[2]{Xu Tan}
\author[5]{Zheqi Dai}
\author[5]{Qiuqiang Kong}
\author[1]{Jianyi Chen}
\author[1]{Jiahao Pan}
\author[1]{Qifeng Liu}
\author[1,*]{Yike Guo}
\author[1,*]{Wei Xue}

\affil[1]{Hong Kong University of Science and Technology}
\affil[2]{Microsoft}
\affil[3]{University of Science and Technology Beijing}
\affil[4]{Hong Kong Baptist University}
\affil[5]{Chinese University of Hong Kong}

\begin{document}

\maketitle

\begin{abstract}

Recent advancements in audio generation have been significantly propelled by the capabilities of Large Language Models (LLMs). The existing research on audio LLM has primarily focused on enhancing the architecture and scale of audio language models, as well as leveraging larger datasets, and generally, acoustic codecs, such as EnCodec, are used for audio tokenization. However, these codecs were originally designed for audio compression, which may lead to suboptimal performance in the context of audio LLM. Our research aims to address the shortcomings of current audio LLM codecs, particularly their challenges in maintaining semantic integrity in generated audio. For instance, existing methods like VALL-E, which condition acoustic token generation on text transcriptions, often suffer from content inaccuracies and elevated word error rates (WER) due to semantic misinterpretations of acoustic tokens, resulting in word skipping and errors. To overcome these issues, we propose a straightforward yet effective approach called X-Codec. X-Codec incorporates semantic features from a pre-trained semantic encoder before the Residual Vector Quantization (RVQ) stage and introduces a semantic reconstruction loss after RVQ. By enhancing the semantic ability of the codec, X-Codec significantly reduces WER in speech synthesis tasks and extends these benefits to non-speech applications, including music and sound generation. Our experiments in text-to-speech, music continuation, and text-to-sound tasks demonstrate that integrating semantic information substantially improves the overall performance of language models in audio generation. 
Our code and demo are available 
\footnote{\begin{minipage}[t]{\linewidth} Demo: https://x-codec-audio.github.io \\ Code: https://github.com/zhenye234/xcodec \end{minipage}}

\end{abstract}

% 两阶段pipeline complex
% 两阶段错误传递 
% 两阶段彼此限制  
% 两阶段过程复杂
% 想实现 GPT-style autoregressive methods 
% In contrast to the remarkable achievements of LLMs, the power of autoregressive models in audio appears to be somewhat locked
% 之前的token 没有一个统一的理解，探索前后逻辑的能力被挖掘，gpt2-like transformer
% 忽略了 semantic ， 之前acoustic预测，需要模型容量去理解acoustic， 
% 主要目的，结合了两者优势，比之前要好。
% why gpt-style 1 大量数据 pretrain 2 简单 灵活性 3，bert maybe two stage 模型应用受限
% 之前两阶段，训练目标，训练过程，结构都不一样，复杂。  错误传递 
% 之前纯acoustic token性能受限
% unlock gpt-style autoregressive methods 
%soundstream 改encodec

\section{Introduction}
 
In recent years, Large Language Models (LLMs) such as GPT \cite{brown2020language} have demonstrated remarkable capabilities in modeling complex, high-dimensional data across various domains, including text and image generation \cite{zhao2023survey,liu2024visual}. Inspired by these successes, there has been significant interest \cite{agostinelli2023musiclm,borsos2023audiolm,wang2023neural,yang2023uniaudio} in exploring the application of LLMs to audio generation.

Audio codecs \cite{zeghidour2021soundstream} have emerged as a critical technique for audio LLMs, bridging the gap between continuous audio waveforms and token-based language models. By discretizing high-rate audio signals into a finite set of tokens, these codecs enable the application of LLM architectures to audio data, leveraging the successes of textual LLMs. 

However, prior research on audio codecs has primarily focused on achieving lower compression rates and higher reconstruction quality \cite{kumar2024high,defossez2022high,yang2023hifi}. Meanwhile, many efforts in audio generation have concentrated on enhancing model architecture, scaling, or leveraging larger datasets. For instance, AudioLM \cite{borsos2023audiolm} adopts a two-stage pipeline that models the acoustic token in an autoregressive way conditioned on the semantic token. VALL-E \cite{wang2023neural}, the first TTS framework to leverage large, diverse, and multi-speaker speech data, demonstrates strong in-context learning capabilities similar to GPT-3, treating TTS as a language modeling task on audio codecs. MusicGen \cite{copet2024simple} generates music using a single-stage transformer LM alongside efficient token interleaving patterns. Similarly, UniAudio \cite{yang2023uniaudio} scaled up to 165K hours of audio and 1B parameters, utilizing LLM techniques to generate tokens for various types of audio, including speech, sounds, music, and singing, given different input conditions.

While these works have shown success in developing audio language models, they all rely on the acoustic codecs such as Encodec \cite{defossez2022high} or Soundstream \cite{zeghidour2021soundstream} for audio tokenization and de-tokenization. However, these acoustic codecs were originally designed for audio compression rather than for audio language models. This misalignment means the design may not be optimal for audio language modeling.

To design a better audio codec for Audio LLMs, we drew inspiration from the initial purpose of LLMs such as GPT, which were designed to process text. These models focus on understanding and generating natural language, which is inherently rich in semantics. Motivated by this, we assume that a better audio tokenizer should encapsulate rich semantic information to facilitate an easy understanding of audio content, thus reducing the language model's burden in interpreting tokens. However, most audio codecs focus on acoustic reconstruction which ignores the semantic information. As a result, LLM essentially tries to predict the local fluctuations of the audio signal, which is difficult, and methods like VALL-E, which condition acoustic token generation on text transcriptions, frequently result in content inaccuracies causing elevated word error rates (WER), stemming from the semantic misinterpretations of acoustic tokens, leading to word skipping and errors.

To address this issue, approaches like SpeechTokenizer \cite{zhang2023speechtokenizer} have attempted to disentangle speech into separate tokens for content and timbre and perform distillation-based semantic and acoustic integration. However, this method may not integrate smoothly with all audio LLMs, especially those requiring uniform token treatment across different layers, such as utilizing flattened codec tokens \cite{yang2023uniaudio,copet2024simple}. 

In this paper, We propose a straightforward yet effective method termed ``X-codec'', which integrates both semantic and acoustic features into a unified tokenization framework. The X-Codec architecture employs a distinctive ``X-shaped'' structure, characterized by two inputs and two outputs, unifying semantic and acoustic information within a single Residual Vector Quantizer (RVQ) structure. This design enables simultaneous embedding learning of semantic richness and acoustic fidelity for every token, resulting in better performance for audio LLM. 

We have conducted comprehensive evaluations of X-Codec across various applications, including text-to-speech, music continuation, and text-to-sound synthesis. The results consistently demonstrate the effectiveness of the proposed method. Furthermore, our comparative evaluation on VALL-E based TTS demonstrates that X-Codec outperforms existing disentanglement techniques, thereby highlighting its efficacy and versatility in advancing audio LLM technologies.

\section{Related Works}

\subsection{Audio Language Model}

The success of Large Language Models (LLMs) has sparked a significant trend in leveraging language foundation models for audio generation tasks \cite{rubenstein2023audiopalm,zhang2024speechlm,wu2023decoder,wu2023speechgen,yang2023uniaudio,chen2023lauragpt}. Audio, much like language, consists of variable-length sequences, making it well-suited for modeling with language foundation models. One pioneering method, AudioLM \cite{borsos2023audiolm}, employs a multi-stage strategy to harness the predictive capabilities of foundation models for generating tokens unconditionally. This approach involves predicting semantic tokens from various conditions (e.g., phonemes, text descriptions, MIDI) in the initial stage, followed by transforming them into acoustic tokens through coarse-to-fine modeling, ultimately generating the waveform. Representative systems such as SPEAR-TTS \cite{kharitonov2023speak} for speech synthesis and MusicLM \cite{agostinelli2023musiclm} for music generation have also been proposed. However, the two-stage process can lead to complexity in training and suboptimal performance due to the separate development of semantic and acoustic tokens, leading to error accumulation.

Conversely, recent advancements have shown that methods employing a single-stage language model outperform two-stage approaches. For example, VALL-E \cite{wang2023neural} utilizes an autoregressive (AR) model to predict the first token and a non-autoregressive (NAR) model to estimate the residual tokens, demonstrating superior performance compared to AudioLM. Similarly, MusicGen \cite{copet2024simple} employs a single-stage transformer language model and incorporates a delay pattern strategy for efficient token interleaving, achieving better results than MusicLM. Other notable works include CLAM-TTS \cite{kim2024clam}, VoiceCraft \cite{peng2024voicecraft}, and UniAudio \cite{yang2023uniaudio}.

Despite recent advancements, directly modeling the intricate low-level acoustic fluctuations with an LLM poses challenges. LLMs are primarily designed for processing natural language, which is inherently rich in semantics. In order to overcome this limitation, we propose X-Codec, a novel enhancement that aims to enrich semantic processing within acoustic codecs. By doing so, we aim to improve the overall performance of audio LLMs.

\begin{figure*}[h] 
  \centering
  \includegraphics[scale=0.60]{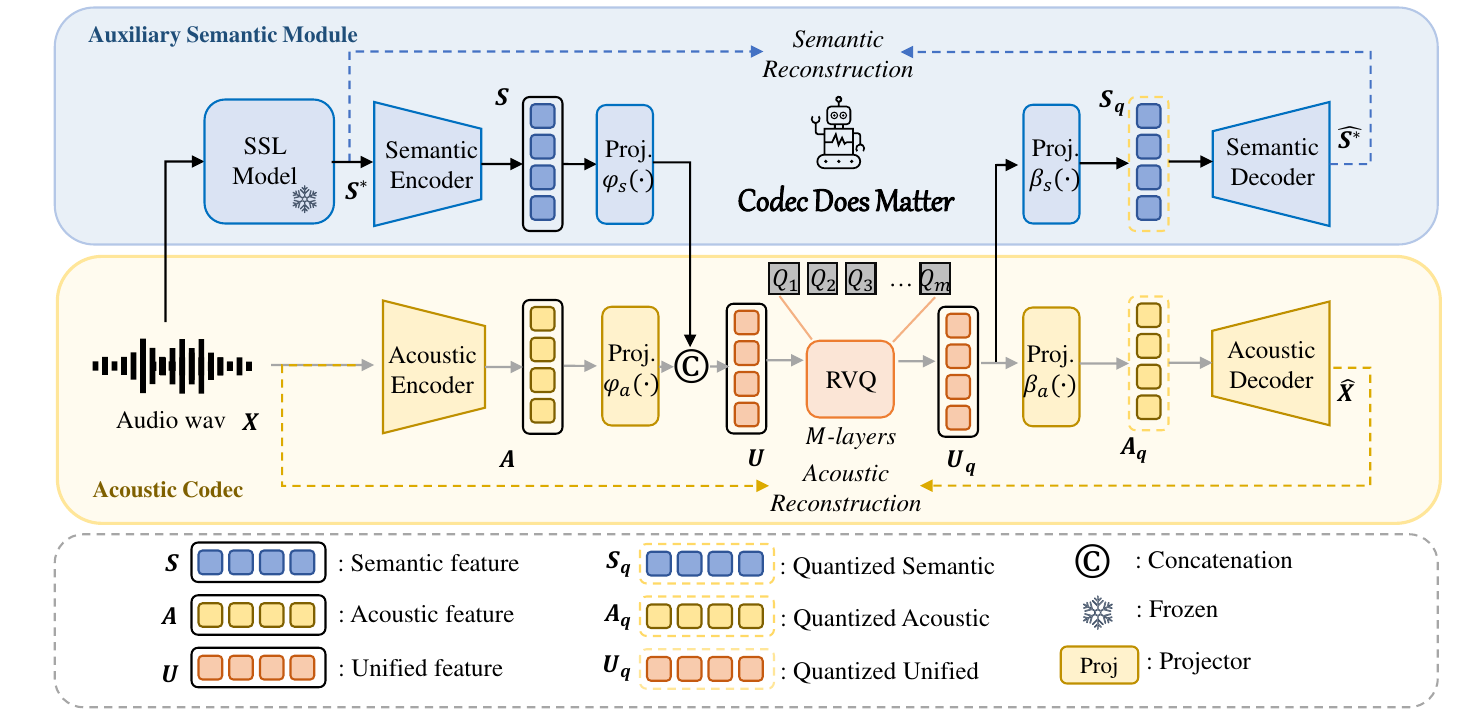}
  \caption{The pipeline of X-codec.}
  \label{fig_xcodec}
\end{figure*}

\subsection{Audio Codec}

Recent advancements have seen a surge in deep learning methodologies employing vector quantization \cite{van2017neural} to reconstruct continuous signals into discrete representations for AR generation. Notably, audio codecs based on the VQ-GAN framework \cite{esser2021taming} have gained prominence. For example, SoundStream \cite{zeghidour2021soundstream} introduces a versatile codec adaptable to various audio types, integrating Residual Vector Quantization (RVQ) and Generative Adversarial Network (GAN) to refine quantization and reconstruction. Similarly, Encodec \cite{defossez2022high} enhances compression through a multi-scale discriminator and a loss-balancing strategy alongside a language model. HiFi-Codec \cite{yang2023hifi} employs Group-Residual Vector Quantization (GRVQ) to minimize the need for extensive codebooks while maintaining high reconstruction fidelity. DAC \cite{kumar2024high} addresses codebook collapse, where some codes remain unused, by applying improved codebook learning to achieve higher compression rates.

These codecs primarily focus on acoustic reconstruction and higher compression rates, often overlooking their potential as tokenizers for audio LLMs. Some attempts have been made to develop more suitable tokenizers for audio LLMs. For example, SpeechTokenizer \cite{zhang2023speechtokenizer} utilizes HuBERT to separate speech into distinct VQ components for content and timbre/acoustic details. This separation improves the modeling of content in the AR stage of VALL-E, while the NAR stage enriches the acoustic details. However, a distillation framework is exploited, this makes SpeechTokenizer may not be compatible with all LLM architectures, especially those that require uniform treatment of tokens, such as methods using flattened codec tokens \cite{yang2023uniaudio,copet2024simple}. Another attempt is presented by SemantiCodec \cite{liu2024semanticodec}, which employs a pre-trained AudioMAE \cite{huang2022masked} to generate distinct semantic and acoustic tokens from mel-spectrograms. However, this method inherits the issues of SpeechTokenizer and introduces additional complexity in token modeling. Moreover, since the audioMAE is performed on 2D time-frequency mel-spectrograms, LLMs must effectively handle dual scales (time and frequency), which may require significant modifications to existing LLM structures.

In contrast, our proposed X-Codec provides a uniform and comprehensive enhancement of semantic information for all tokens, resulting in significant performance improvements for existing audio LLMs without requiring any structural modifications.

\section{Methods}

In this section, we propose X-codec, a straightforward yet effective method to overcome the semantic shortcomings of the current acoustic codecs. 

\subsection{Acoustic Audio codec}

As illustrated in Figure \ref{fig_xcodec}, our model builds upon the framework established by existing acoustic codecs such as Encodec \cite{defossez2022high} and DAC\cite{kumar2024high}. An acoustic audio codec is composed of three main components: an acoustic encoder, a quantizer, and an acoustic decoder. The input of the codec is the raw waveform $\textbf{X} \in \mathbb{R}^{n}$, where $n$ represents the number of waveform samples. This waveform is fed into the acoustic encoder, which consists of several convolutional layers and employs temporal downscaling to extract frame-level latent acoustic features $\textbf{A} \in \mathbb{R}^{H_a \times T}$, where $H_a$ denotes the hidden size of the acoustic features and $T$ is the number of frames. These continuous features are then transformed into a series of discrete tokens $\textbf{Q} \in \mathbb{R}^{M \times T}$ using a Residual Vector Quantizer (RVQ) with $M$ quantizer layers. During training, a specific codebook for the quantizer is learned, enabling the conversion of discrete tokens back to continuous features $\textbf{A}_q \in \mathbb{R}^{H_a \times T}$. The acoustic decoder then reconstructs the waveform $\hat{\textbf{X}}$ from $\textbf{A}_q$ using several convolutional layers and temporal upsampling. The training process is supervised using various losses, including mel loss, STFT loss, and GAN loss, to ensure high-quality acoustic reconstruction.

\subsection{Analysing Semantic Shortcoming}

In this section, we investigate the impact of acoustic codecs on the performance of audio LLMs, focusing specifically on VALL-E, a pioneering model that leverages language model principles for text-to-speech. Our analysis reveals that training VALL-E using Encodec results in high word error rates (WER) and frequent inaccuracies in content generation. For example, when the input text ``he passed through Henley Saint Albans and came so near to London as Harrow on the Hill'' is synthesized, it is erroneously produced as ``he passed through henley saint albeans and camsel knew to lunglan as herold the lor''. This misinterpretation, which is beyond simply improving the audio quality, suggests a fundamental limitation in Encodec's ability to differentiate phonemes, possibly due to its inadequate semantic processing capabilities.

To substantiate the above hypothesis, we conducted Phonetic Discriminability ABX Tests to evaluate the phonetic discriminability of Encodec's representations. The details are provided in the experiment section. Our findings reveal that Encodec's representations exhibit poor phonetic discriminability, which confirms the presence of semantic inadequacies in the codec. Based on these results, we assert that these semantic shortcomings are a significant contributing factor to the observed inaccuracies of language model based audio generation.

To effectively address these semantic limitations, we introduce a novel approach that integrates more comprehensive semantic features into the codec's architecture. This enhancement is designed to enrich the codec's understanding of audio content, thereby alleviating the interpreting load on the language model. Detailed elaboration of this method is provided in the subsequent section.

\subsection{Designing Auxiliary Semantic Module}

Our approach employs a straightforward method that enhances audio codecs by directly concatenating semantic and acoustic features. Initially, we extract the semantic feature vector $\textbf{S}^* \in \mathbb{R}^{H_s \times T}$ from the audio waveform $\textbf{x}$. This extraction utilizes a self-supervised, pre-trained model such as HuBERT \cite{hsu2021hubert} or wav2vec 2.0 \cite{baevski2020wav2vec}. The extracted features are then processed through multiple convolutional layers within a semantic encoder to yield the refined semantic feature vector $\textbf{S}$. Concurrently, the acoustic branch produces the feature $\textbf{A}$. These outputs, $\textbf{S}$ and $\textbf{A}$, are subsequently concatenated using a linear projection $\phi$, formulated as:

\begin{equation}
\textbf{U} =  concat(\phi_s(\textbf{S}), \phi_a(\textbf{A})) ,
\end{equation}
where the concatenated feature $\textbf{U} \in \mathbb{R}^{H_u \times T}$ is designed to maximize information preservation from both semantic and acoustic sources. This combined feature is then subject to RVQ using an $M$-layer quantizer, resulting in tokens that encapsulate a rich mixture of semantic and acoustic information.

The quantized feature $\textbf{U}_q$ is designed to meet the decoder's objectives through two projectors, $\beta_s$ and $\beta_a$, which enable the decoders to reconstruct the original semantic feature $\hat{\textbf{S}}^*$ and the audio waveform $\hat{\textbf{x}}$. We adhere to established acoustic reconstruction methods from previous works while introducing a Mean Squared Error (MSE) loss specifically for the reconstruction of semantic features. Furthermore, a constant weight $\gamma$ is applied to the semantic loss to ensure that its scale is aligned with other losses, thus promoting a balanced training objective.
\section{Experiments}

Given that established audio codecs such as Encodec, Speechtokenizer, and DAC are trained on diverse datasets with varying configurations, we meticulously design experiments to rigorously evaluate the efficacy of our proposed solution, X-Codec. To ensure a fair and unbiased comparison, each experiment employs a baseline acoustic codec that is precisely aligned with our X-Codec in terms of training data, training steps, and other hyperparameters. The primary distinction between the baseline codec and X-Codec lies in the exclusion of the auxiliary semantic module in the baseline configuration. This controlled experimental design enables us to isolate and evaluate the specific contributions of our semantic enhancements to the overall performance of the audio LLMs.

\subsection{Text-to-Speech}

In this subsection, we critically evaluate the performance of various audio codecs in training the VALL-E model for zero-shot Text-to-Speech (TTS) tasks. Our investigation is guided by two primary objectives:
\begin{itemize}
\item To determine whether the X-Codec can enhance the performance of audio LLMs in TTS applications.
\item To evaluate the comparative advantages of X-Codec over the disentanglement strategy employed by SpeechTokenizer, specifically within the context of the VALL-E model.
\end{itemize}

\subsubsection{Baselines}

For a comprehensive comparison, we employ several state-of-the-art neural audio codecs as baselines:

\begin{itemize}
\item \textbf{EnCodec} \footnote{\url{https://huggingface.co/facebook/encodec\_24khz}}: The open-source EnCodec model \cite{defossez2022high}, trained on a diverse range of 24kHz audio data, can compress audio to bitrates between 1.5 and 24.0 kbps while maintaining high fidelity.
\item \textbf{DAC} \footnote{\url{https://github.com/descriptinc/descript-audio-codec}}: The open-source DAC model \cite{kumar2024high} utilizes enhanced VQ techniques. For our experiments, we employ the official 16kHz version.
\item \textbf{SpeechTokenizer} \footnote{\url{https://github.com/ZhangXInFD/SpeechTokenizer}}: This model \cite{zhang2023speechtokenizer} is a unified speech tokenizer that leverages distinct VQ layers to separate speech into content and timbre components. We utilize their official checkpoints in our evaluations.
\end{itemize}
 
\subsubsection{Training Details of X-Codec}
Given our objective to assess the efficacy of X-Codec in leveraging semantic information, we meticulously align our experimental setup with that used for SpeechTokenizer. Both models are trained on the same dataset, LibriSpeech, and utilize the same pre-trained self-supervised representations from HuBERT-base-ls960 \footnote{\url{https://huggingface.co/facebook/hubert-base-ls960}}. To ensure comparability, we also adopt the strategy of employing the average representation across various layers of HuBERT as our semantic training objective.

\subsubsection{Training Details of VALL-E}
For reproduction of the VALL-E, we utilize the resources specified in the provided repository \footnote{https://github.com/lifeiteng/vall-e}. The training data is the LibriTTS, retaining the default settings as specified in the repository, except for the learning rate during the AR stage, which is adjusted to 0.01 to enhance model stability. The training process span 100 epochs for the AR stage and 200 epochs for the non-autoregressive (NAR) stage, same for all audio codecs for a fair comparison.

\subsubsection{Evaluation Metrics}
To assess the performances of zero-shot TTS systems, we employ the following metrics:
\begin{itemize}
    \item \textbf{WER (Word Error Rate)}: We utilize an Automatic Speech Recognition (ASR) model to transcribe the generated audio \cite{wang2023neural}. The discrepancies between these transcriptions and the original texts are quantified using WER, providing a critical measure of audio intelligibility.
    
    \item \textbf{Sim-O (Similarity Objective)}: This metric assesses the objective similarity between synthesized speech and the original reference speech. Sim-O uses feature embeddings extracted from a pre-trained speaker verification model to measure this similarity \cite{hsu2021hubert,kim2024clam}\footnote{\url{https://github.com/microsoft/UniSpeech/tree/main/downstreams/speaker_verification}}, reflecting the codec's ability to preserve speaker characteristics.
    
    \item \textbf{UTMOS}: We evaluate the audio quality using UTMOS, a Speech MOS (Mean Opinion Score) predictor \cite{saeki2022utmos}\footnote{\url{https://github.com/tarepan/SpeechMOS}} that automatically measures the naturalness of speech. This metric provides insights into the overall auditory quality of the synthesized speech.
\end{itemize}

% \begin{table*}[h]
%   \centering
%     \caption{Objective performance comparison on \textit{continuation} zero-shot speech
% synthesis tasks using VALL-E trained on LibriTTS  \textbf{with different audio codec}. Abbreviation: C (Common Voice), DNS (DNS Challenge 4 speech), AS (AudioSet), FSD (FSD50K), J (Jamendo), V (VCTK), M(MUSDB) }
% \resizebox{0.9\textwidth}{!}{%
%   \begin{tabular}{lccccccc}
%     \toprule
%     Codec   & Training Data of   &  \multicolumn{3}{c}{\textbf{VALL-E AR stage}} & \multicolumn{3}{c}{\textbf{VALL-E AR+NAR stages}}   \\
%     & Audio Codec  &WER $\downarrow$  &SIM-O $\uparrow$  &UTMOS $\uparrow$      &WER $\downarrow$  &SIM-O $\uparrow$  &UTMOS $\uparrow$ \\
%     \midrule
%     GT & - & 2.23 &  0.67 &  4.10 &  2.23 &  0.67 &  4.10\\
%     \midrule
%    Encodec \cite{defossez2022high} & C+DNS+AS+FSD+J& 47.17  &  0.09 &  1.24 &  6.37 &  0.33 &  3.02\\
%    DAC \cite{kumar2024high}  &  C+DNS+V+AS+J+M & 85.55 &  0.03 &  1.24 &  6.81 &  0.34 &  3.31 \\
%    Speechtokenizer \cite{zhang2023speechtokenizer} & LibriSpeech & 7.53  &  0.10 &  1.26 &  5.24 &  0.36 & 3.84 \\
%     \bottomrule
%    Baseline Acoustic Codec & LibriSpeech & 22.32   & 0.16 & 3.15   & 7.70   & 0.41 & 3.89 \\
%     X-Codec & LibriSpeech &  5.27 &  0.22 &  3.85 &  4.07 &  0.42 & 4.16   \\
%     \bottomrule
%   \end{tabular}
%   }
%   \label{valle_exp}
% \end{table*}

\begin{table*}[h]
  \centering
    \caption{Objective performance comparison on \textit{continuation} zero-shot speech
synthesis tasks using VALL-E trained on LibriTTS  \textbf{with different audio codec}. Abbreviation: C (Common Voice), DNS (DNS Challenge 4 speech), AS (AudioSet), FSD (FSD50K), J (Jamendo), V (VCTK), M(MUSDB) }
\resizebox{0.9\textwidth}{!}{%
  \begin{tabular}{lccccccc}
    \toprule
    Codec   & Training Data of   &  \multicolumn{3}{c}{\textbf{VALL-E AR stage}} & \multicolumn{3}{c}{\textbf{VALL-E AR+NAR stages}}   \\
    & Audio Codec  &WER $\downarrow$  &SIM-O $\uparrow$  &UTMOS $\uparrow$      &WER $\downarrow$  &SIM-O $\uparrow$  &UTMOS $\uparrow$ \\
    \midrule
    GT & - & 2.23 &  0.67 &  4.10 &  2.23 &  0.67 &  4.10\\
    \midrule
   Encodec \cite{defossez2022high} & C+DNS+AS+FSD+J& 47.17  &  0.09 &  1.24 &  6.37 &  0.33 &  3.02\\
   DAC \cite{kumar2024high}  &  C+DNS+V+AS+J+M & 85.55 &  0.03 &  1.24 &  6.81 &  0.34 &  3.31 \\
   Speechtokenizer \cite{zhang2023speechtokenizer} & LibriSpeech & 7.53  &  0.10 &  1.26 &  5.24 &  0.36 & 3.84 \\
    \bottomrule
   Baseline Acoustic Codec & LibriSpeech & 22.32   & 0.16 & 3.15   & 7.70   & 0.41 & 3.89 \\
    X-Codec-hubert & LibriSpeech &  5.27 &  0.22 &  3.85 &  4.07 &  0.42 & 4.16   \\
    X-Codec-wavlm-base-plus & MLS English &  4.83 &  0.24 &  4.02 &  3.26 &  0.41 & 4.22   \\
    \bottomrule
  \end{tabular}
  }
  \label{valle_exp}
\end{table*}

\subsubsection{Zero-shot TTS Results}
We use librispeech-test-clean \cite{panayotov2015librispeech}for zero-shot TTS evaluation following VALL-E-continual-setting \cite{wang2023neural}. The results in Table \ref{valle_exp} demonstrate the following key findings:

\begin{itemize}
\item When comparing both X-Codec and SpeechTokenizer against the baseline and other acoustic codecs like DAC and Encodec, we observe improvements in WER. This supports our hypothesis that integrating semantic information helps audio LLMs better understand content.
\item Comparing the baseline acoustic codec and SpeechTokenizer, SpeechTokenizer exhibited lower Sim-O scores. We attribute this reduction to its initial disentanglement phase, which exclusively focuses on content prediction. This specialization potentially hampers the NAR phase's ability to accurately reconstruct speaker timbre when conditioned solely on tokens derived from the primary content-focused stage, resulting in poor speaker similarity.
\item  X-Codec not only shows better WER but also higher Sim-O and UTMOS scores compared to SpeechTokenizer. This confirms the effectiveness of our approach, indicating that our codec handles the integration of semantic and acoustic information more proficiently.
\end{itemize}

\begin{table}
\centering
\resizebox{0.4\textwidth}{!}{%
\begin{tabular}{lccc}
\hline
\textbf{Model} & \textbf{$n_q$} & \textbf{ within $\downarrow$} & \textbf{ across $\downarrow$} \\
\hline
hubert-ls-960 \cite{hsu2021hubert} & - & 3.3 & 4.1 \\
Encodec \cite{defossez2022high}& 1 & 21.5 & 28.3 \\
Encodec \cite{defossez2022high} & 8 & 17.5 & 27.0 \\
DAC \cite{kumar2024high} & 1 & 26.3 & 32.7 \\
DAC \cite{kumar2024high}& 12 & 21.7 & 33.2 \\
Speechtoknizer \cite{zhang2023speechtokenizer}& 1 & 3.5  & 4.3 \\
Speechtoknizer\cite{zhang2023speechtokenizer}& 8 & 3.6  &  4.5 \\
    \midrule
Baseline Acoustic Codec & 1 & 26.4 & 31.2 \\
Baseline Acoustic Codec & 8 & 20.1 & 28.3 \\
X-Codec & 1 & 3.9 & 4.9 \\
X-Codec & 8 & 3.3 & 4.3 \\
\hline
\end{tabular}
}
\caption{Comparison of Phonetic Discriminability within and across  ABX error rate for various models, with different $n_q$ values. Lower values indicate better performance.}
\label{abx}
\end{table}

\subsubsection{Analysing the Effect of Codec}
To further analyse the above results caused by different audio codecs, we evaluate phonetic discriminability using the ABX error rate \cite{schatz2013evaluating}. This metric assesses how well different codecs can distinguish between similar phonetic sounds within and across various contexts. We specifically examine the continuous representations for VQ as indicated by the results in the following table \ref{abx}. We compare the performance of various models in terms of within and across phonetic discriminability:

Key insights include:
\begin{itemize}
\item Both SpeechTokenizer and X-Codec significantly outperform pure acoustic codecs like Encodec and DAC in phonetic discriminability, which supports our claim that enhancing semantic understanding in codecs helps modelling content such as phonetic details.
\item The X-Codec demonstrates a notable trend of improved phonetic discriminability with an increase in the number of quantizations (nq). Specifically, as nq increases from 1 to 8, the ABX error rates consistently decrease, thereby highlighting effectiveness of the X-Codec's design in enhancing semantic integration across multiple quantization layers.

\item In contrast, the SpeechTokenizer, while exhibiting commendable performance at a lower quantization level (nq = 1), fails to show significant improvement as nq is increased. This suggests a design limitation; the codec's reliance on the initial quantization to carry semantic information restricts its ability to process a broader spectrum of semantic information. Notably, the performance of X-Codec at nq = 8 significantly exceeds that of SpeechTokenizer.
\end{itemize}

These results underline the effectiveness of our method in facilitating enhanced semantic integration, leading to better phonetic discriminability and audio LLMs. In addition, these results also show that our simple concatenate methods surpass disentangle methods such as speechtokenizer.

\subsection{Music and Sound Generation}

To the best of our knowledge, this is the first exploration into the potential benefits of incorporating semantic information into audio codecs for enhancing music and general sound generation through audio LLMs. Conventional methods for general audio representation learning, aiming at capturing the semantic discriminability of audios, are generally based on 2D mel-spectrogram, such as AudioMAE \cite{huang2022masked} and Beats \cite{chen2022beats}. These methods are in stark contrast to traditional codecs that process audio sequentially, frame-by-frame. This difference poses challenges for direct integration into existing audio generation frameworks.

To bridge this gap, we have developed a variant of HuBERT, specifically adapted for general audio, which we refer to as HuBERT-General-Audio. This HuBERT-General-Audio is trained on an expansive internal dataset of approximately 200,000 hours, with a similar distribution as AudioSet. Additionally, our proposed X-Codec is also trained using these data for 400,000 steps until convergence, incorporating the HuBERT-General-Audio model within its semantic module. For a fair comparison, we train a baseline acoustic codec under identical settings but excluding semantic information. 

\subsubsection{Training Details of Self-Supervised General Audio Representation}
HuBERT-General-Audio is trained using 8 NVIDIA H800 GPUs on 2.6 million tokens across 325,000 iterations. For training stability, we adopt an inverse square root learning schedule, a modification from the polynomial decay schedule originally utilized in \cite{hsu2021hubert}. The learning rate is set at 0.0003 with warmup steps of 32,000. Unlike the original HuBERT, which utilizes MFCCs as the training target unit designed specifically for speech, our model leverages the first VQ layer of Encodec as the training target for acoustic unit discovery in the general audio. This choice eliminates the need for the K-means discretization step, saving significant time and computational resources.
 
\begin{table}[t!]
\centering
\resizebox{0.4\textwidth}{!}{%
\begin{tabular}{lccc}
\hline
\textbf{Model} & \textbf{FD} $\downarrow$ & \textbf{FAD} $\downarrow$ & \textbf{FD-MERT-layer-9} $\downarrow$ \\
\hline
Acoustic codec   & 16.17 & 1.43  & 2.88 \\
X-Codec & 12.66 & 1.37 & 2.62  \\
\hline
\end{tabular}
}
\caption{Comparison between baseline acoustic codec and our X-Codec on music continue.}
\label{music}
\end{table}

\subsubsection{Music Continuation}

{\emph{Training Details}}: 
Acquiring high-quality text-music pair data is challenging; therefore, we gathered approximately 100,000 hours of music-only data, including about one million songs for the music continuation task. We deployed nanoGPT \footnote{https://github.com/karpathy/nanoGPT} to implement a GPT-2-medium (approximately 300M parameters) \cite{radford2019language} as our generative model. This model utilizes the first  VQ  from our codec to construct the training sequences, with additional experiments involving multiple VQs detailed in the appendix. We set the block size of sequence modelling to 4096, corresponding to roughly 82 seconds of audio, and adjust the vocabulary size from 50,257 to 1024, matching our codec's codebook size. Other training hyperparameters are consistent with previous GPT-2-medium configurations. We train 300,000 steps on 8 NVIDIA H800 GPUs. The batch size is set to 20, with a learning rate of 3e-4 and a warmup phase of 2000 steps.

\emph{Experiments}:
For music continuation, we randomly crop 600 samples with each 40 seconds in duration from the MUSDB18 dataset \cite{rafii2017musdb18}. The initial 10 seconds of each sample are used as prompts for the audio LLM, while the subsequent 30 seconds are generated by the model. These generated segments are then compared against the corresponding ground truth (GT) segments. To ensure that the assessment is independent of the codec's reconstruction fidelity, both the generated and GT audio are reconstructed using the first VQ layer of the codec, ensuring performance differences attributed solely to the generative models themselves.

The evaluation metrics of the generated music include: Frechet Distance (FD) computed using features from Pretrained Audio Neural Networks (PANNs) \cite{kong2020panns}, Frechet Audio Distance (FAD), and FD-MERT Layer 9 \cite{li2023mert}. The results, as summarized in Table \ref{music}, reveal that the X-Codec significantly outperforms the baseline acoustic codec across all metrics. This superior performance indicates the X-Codec has a better understanding and enabling more effective reproduction of complex musical structures.

\subsubsection{Text-to-Sound}
\begin{table}[t]
  \centering
\resizebox{0.35\textwidth}{!}{%
\begin{tabular}{lcccc}
\hline
\multicolumn{1}{c}{\textbf{Model}} & \textbf{FD}$\downarrow$ & \textbf{IS}$\uparrow$ & \textbf{FAD}$\downarrow$ & \textbf{CLAP}$\uparrow$ \\ \hline
Acoustic codec                   & 59.03          & 3.89         & 6.19            & 0.417          \\
X-Codec                       & 46.31          & 5.29         & 4.10            & 0.483          \\ \hline
\end{tabular}
}

\caption{Comparison between baseline acoustic codec and X-Codec on text-to-sound tasks.}
\label{t2s}
\end{table}

\begin{table*}[t]
\centering
\resizebox{\textwidth}{!}{%
\begin{tabular}{lcccccccccccc}
\hline
Model/Datasets & ESC-50 & US8K & FSD50K & VIVAE & FMA & MTT & IRMAS & MS-DB & RAVDESS & A-MNIST & SLURP & EMOVO \\
\hline
DAC \cite{kumar2024high}  & 27.65 & 45.16 & 7.08 & 30.80 & 38.50 & 27.69 & 30.33 & 51.79 & 37.50 & 73.59 & 7.72 & 23.46 \\
Encodec \cite{defossez2022high} & 30.60 & 64.47 & 8.63 & 31.59 & 39.5 & 26.57 & 28.16 & 64.47 & 31.60 & 78.71 & 8.44 & 25.51 \\
\hline

Baseline Acoustic Codec & 40.00 & 55.38 & 11.10 & 37.70 & 48.75 & 32.26 & 36.26 & 62.77 & 47.22 & 82.58 & 9.28 & 24.83 \\
Hubert-general-audio & 69.95 & 74.87 & 34.27 & 48.35 & 64.50 & 43.35 & 49.80 & 74.09 & 69.10 & 99.43 & 21.27 & 35.03 \\
X-Codec & 69.85 & 75.37 & 34.05 & 49.40 & 64.63 & 42.95 & 52.24 & 75.63 & 68.40 & 99.48 & 21.65 & 35.20 \\
\hline
\end{tabular}%
}
\caption{Performance of semantic representation on the ARCH benchmark. The table shows the performance of various models across different domains. ESC-50 \cite{ESC50}, US8K \cite{US8K}, FSD50K \cite{FSD50K}, and VIVAE \cite{VIVAE} represent performance on Acoustic Events. FMA \cite{FMA}, MTT \cite{MTT}, IRMAS \cite{IRMAS}, and MS-DB \cite{medleydb} indicate performance in the Music domain. RAVDESS \cite{RAVDESS}, AudioMNIST \cite{AudioMNIST}, SLURP \cite{SLURP}, and EMOVO \cite{EMOVO} reflect performance in the Speech domain. Higher values indicate better performance across these tasks.}
\label{arch}
\end{table*}

{\emph{Training Details}}: 
Still, GPT-2-medium (approximately 300M parameters) are adopted for conditional text-to-sound tasks, where the condition embedding is extracted from text captions using LAION-CLAP \cite{wu2023large} and linearly projected from 512 dimensions to 1024 dimensions for GPT input. The training data consists of approximately 400 hours of audio content sourced from the AudioCaps dataset \cite{kim2019audiocaps} and the AudioSet SL subset from the WavsCaps dataset \cite{mei2024wavcaps}. All audio samples are uniformly resampled to a 16kHz sampling rate. The first four tokens from the VQ layers are preprocessed and flattened to configure the GPT model's block size to 2000, corresponding to a processing rate of 50Hz. The training process spans 80,000 steps on four NVIDIA 4090 GPUs, with a batch size of 8 and a learning rate of 3e-4. A warmup phase of 2000 steps is employed to optimize the training process.

{\emph{Evaluation Metrics}}:
following \cite{huang2023make} and \cite{liu2023audioldm}, we calculate Frechet Distance (FD), Inception Score (IS), Frechet Audio Distance (FAD) for text-to-audio generation. In addition, CLAP score \cite{huang2023make} is used to evaluate the correspondence between the generated audio and the text prompt.

{\emph{Experiment Results}}:
As shown in Table \ref{t2s}, the proposed X-Codec significantly outperforms the baseline acoustic codec across all metrics. These results demonstrate that semantic information integration significantly enhances the codec's capability, underscoring the value of semantic enrichment in audio generation tasks.

\subsubsection{Analysing the Effect of Codec}
We hypothesize that the enhanced audio generation capabilities of the audio LLMs are attributed to the improved semantic understanding facilitated by the X-Codec. To validate this hypothesis, we employ the ARCH benchmark \cite{la2024benchmarking} to evaluate the audio semantic understanding, and the benchmark is a comprehensive framework specifically designed to evaluate automatic recognition learning methods across a diverse range of audio classification domains, including acoustic events, music, and speech. The results from this benchmark are shown in Table \ref{arch}.

Our findings indicate that HuBERT-general-audio significantly outperforms traditional acoustic codecs such as DAC, Encodec, and the baseline acoustic codec across all metrics. This improvement highlights the enhanced semantic understanding of X-Codec for general audio, which appears to be lacking in conventional acoustic audio codecs.

Moreover, X-Codec achieves performance that is comparable or even superior to HuBERT-general-audio, confirming the effectiveness of our approach to enhancing semantic processing within codecs. This equivalence or improvement indicates the capability of X-Codec to integrate semantic information robustly.

\subsection{Limitation}
 
While our method significantly enhances the performance of codecs for LLMs by integrating semantic information, it does come with certain trade-offs. According to the principle of "no free lunch," improving one aspect of a system often involves compromises in others. In the case of our enhanced codecs, the primary limitation lies in their potential impact on the original functionality of codecs, which is compression for information transmission. The introduction of a semantic extraction layer adds additional computational overhead, potentially increasing the time required for processing. This can affect the efficiency of the codec when used in applications where rapid data compression and transmission are critical. Consequently, while our approach offers substantial benefits for semantic understanding and audio processing, it may not be as effective in contexts where high-speed data compression is paramount.
\begin{table}[h!]
\centering
\resizebox{0.35\textwidth}{!}{%
\begin{tabular}{lccc}
\hline
\textbf{Model} & \textbf{Mel DT.} & \textbf{STFT DT.} & \textbf{UTMOS} \\
\hline
Baseline ($n_q$=$1$) & 0.79 & 0.73 & 2.96 \\
Baseline ($n_q$=$8$) & 0.54 & 0.58 & 3.72 \\
X-codec ($n_q$=$1$) & 0.86  & 0.77  & 3.71 \\
X-codec ($n_q$=$8$) & 0.62 & 0.63  & 4.01 \\
\hline
\end{tabular}
}
\caption{Comparison of reconstruction based on Mel DT., STFT DT., and UTMOS metrics using 1000 LibriTTS speech samples. “DT.” is short for distance}
\end{table}

Furthermore, the integration of semantic layers can slightly impair certain acoustic metrics such as Mel and STFT distance, which are crucial for maintaining the fidelity of compressed audio. However, it is essential to note that these trade-offs are counterbalanced by significant improvements in human auditory perception, as evidenced by the UTMOS scores.

\section{Conclusion}

In this paper, we introduced X-codec, an advanced audio codec that integrates semantic information through self-supervised learning models to enhance performance in large language models, specifically in text-to-speech synthesis, music continuation, and general audio classification tasks. Our evaluations demonstrate that X-codec significantly improves semantic understanding and audio generation quality across a variety of domains.  

% In this paper, we first introduce the a unified audio tokenization approach that elegantly integrates semantic and acoustic information and can be used in language foundation model in a plug-and-play way. Then, our empirical studies demonstrates enhanced capabilities in both semantic preservation and audio quality, which is beneficial in understanding and generation tasks. Moreover, the exploration of the application potentials of X-Codec in GPT-style audio generation, showing promising and comparative results. With our futher released code, checkpoints or even API, our novel idea and solid tokenization method have great potential to help the community to build the promising and powerful audio foundation model in the future.
\bibliographystyle{unsrtnat}
\bibliography{neurips_2024}

\newpage
% \input{secs/checklist}
% \appendix
% \section{appendix}
\section{Model Details}

Acoustic Encoder: Following the design principles in \cite{kumar2024high}, our encoder comprises four convolutional encoder blocks, each tailored to progressively downsample the input audio waveform at rates [2, 4, 5, 8]. This structured reduction ensures efficient encoding while preserving essential audio characteristics. The final output from the encoder has a hidden size of 256, ensuring detailed feature representation. The total model size for the encoder stands at 12.18MB.

Residual Vector Quantizer (RVQ): A key component in our X-codec, the RVQ utilizes the techniques established by \cite{zeghidour2021soundstream}. We update the codebook entries using an exponential moving average and apply a straight-through estimator to facilitate the gradient computation during backpropagation. To bolster training effectiveness and adaptability, commitment loss is incorporated, and RVQ layers are randomly selected from options [1, 2, 3, 4, 8] during training. This variability allows the model to adapt more dynamically to different audio characteristics.

Acoustic Decoder: The decoder is designed to mirror the encoder’s architecture, featuring four layers that upsample the audio data at inverse rates [8, 5, 4, 2]. This symmetry between the encoder and decoder helps in effectively reconstructing the audio signal from its encoded state. The decoder's model size is approximately 19.27 MB.

Semantic Encoder and Decoder: To further refine the semantic aspects of the audio signals, we incorporate two additional convolutional blocks within both the semantic encoder and decoder, each with a hidden size of 768. This setup enhances the model’s ability to process and integrate semantic information effectively.
\section{Music Continue with 4 VQ Flattern}
% xcodec 
% frechet_audio_distance: 0.2029561
% kullback_leibler_divergence_sigmoid: 0.7827974
% kullback_leibler_divergence_softmax: 0.1475402
% inception_score_mean: 1.5393551
% inception_score_std: 0.0818533
% frechet_distance: 3.4788028

% 0.7972045598772866 

% frechet_audio_distance: 7.0863215
% kullback_leibler_divergence_sigmoid: 0.8706447
% kullback_leibler_divergence_softmax: 0.1659164
% inception_score_mean: 1.6810636
% inception_score_std: 0.1291735
% frechet_distance: 3.8605033

% 1.1233672557857517   
In this section, we expand our music continuation experiments to include conditions with four flattened vector quantizations (VQs) to further validate the effectiveness of our approach. While the experimental details remain consistent with previous setups, the use of four VQs necessitates a shorter segment length of approximately 20 seconds due to increased data density. We prompt the model with 5 seconds of audio and generate  5 seconds, aiming to assess the performance enhancement under these conditions.

\begin{table}[h]
\centering
\caption{Comparison between baseline acoustic codec and our X-codec on music continuation.}
\resizebox{0.4\textwidth}{!}{%
\begin{tabular}{lccc}
\hline
\textbf{Model} & \textbf{FD} $\downarrow$ & \textbf{FAD} $\downarrow$ & \textbf{FD-MERT-layer-9}  $\downarrow$  \\
\hline
Acoustic codec   &3.86  & 7.08 & 1.12  \\
X-codec & 3.47 & 0.20  &0.79   \\
\hline
\end{tabular}
}
\label{music}
\end{table}
The results underscore a noticeable improvement in performance with the X-codec, particularly highlighted by significant enhancements in Frechet Audio Distance (FAD) and perceptual quality metrics, suggesting that our X-codec not only maintains but also amplifies its efficacy in generating musically coherent and contextually rich outputs.

\end{document}